\documentclass[10pt,times]{article}

\usepackage[margin=1in]{geometry}

\usepackage{graphicx}

\graphicspath{{figures_reduced/}}
\usepackage[dvipsnames]{xcolor}
\usepackage{color}
\usepackage{comment}
\usepackage{makecell}

\usepackage{booktabs}

\usepackage{subfigure}
\usepackage{wrapfig}
\usepackage{multirow}
\usepackage{xspace}
\long\def\symbolfootnote[#1]#2{\begingroup%
\def\thefootnote{\fnsymbol{footnote}}\footnote[#1]{#2}\endgroup}
\usepackage{amsmath} 
\usepackage{amsthm}
\usepackage{amsfonts} 
\usepackage{natbib}

\usepackage{hyperref}

\newcommand{\rev}[1]{\textcolor{black}{#1}}
\newcommand{\zoya}[1]{\textcolor{black}{#1}}
\newcommand{\zoyagain}[1]{\textcolor{black}{#1}}
\newcommand{\aaron}[1]{\textcolor{black}{#1}}
\usepackage{authblk}
\makeatletter
\renewcommand\AB@affilsepx{, \protect\Affilfont}
\makeatother

\usepackage{fancyhdr}
\begin{document}

\title{Towards Better User Studies in Computer Graphics and Vision}

\author[1]{Zoya Bylinskii} 
\author[2]{Laura Herman}
\author[1]{Aaron Hertzmann} 
\author[2]{Stefanie Hutka}
\author[4]{Yile Zhang\footnote{Previously intern at Adobe Research}}
\affil[1]{Adobe Research}
\affil[2]{Adobe Inc.}
\affil[4]{University of Washington}

\maketitle

\begin{abstract}
\zoya{Online crowdsourcing platforms have made it increasingly easy to perform evaluations of algorithm outputs with survey questions like ``which image is better, A or B?'', leading to their proliferation in vision and graphics research papers. Results of these studies are often used as quantitative evidence in support of a paper's contributions. \zoyagain{On the one hand we argue that,} when conducted hastily as an afterthought, such studies lead to an increase of uninformative, and,} \aaron{potentially,} \zoya{misleading conclusions.} \zoyagain{On the other hand, in these same communities, user research is underutilized in driving project direction and forecasting user needs and reception.}
\zoya{We call for increased attention to both the design and reporting of user studies in computer vision and graphics papers towards (1)~improved replicability and (2)~improved project direction. Together with this call, we offer an overview of methodologies from user experience research (UXR), human-computer interaction (HCI), and \zoyagain{applied perception} to increase exposure to the available methodologies and best practices. We discuss foundational user research methods (e.g., needfinding) that are presently underutilized in computer vision and graphics research, but can provide valuable project direction.  We provide further pointers to the literature for readers interested in exploring other UXR methodologies. Finally, we describe broader open issues and recommendations for the research community.} %We encourage authors and reviewers alike to recognize when in the project timeline a user study would be most informative.} %, that not every research contribution requires a user study, and that a misguided emphasis on user studies can incentivise perfunctory studies.}
\end{abstract}

% OLD ABSTRACT:
%Online crowdsourcing platforms make it easy to perform evaluations of algorithm outputs with surveys that ask questions like ``which image is better, A or B?'') The proliferation of these ``user studies'' in vision and graphics research papers has led to an increase of hastily conducted studies that are sloppy and uninformative at best, and potentially harmful and misleading. We argue that more attention needs to be paid to both the design and reporting of user studies in computer vision and graphics papers. In an attempt to improve practitioners' knowledge and increase the trustworthiness and replicability of user studies, we provide an overview of methodologies from user experience research (UXR), human-computer interaction (HCI), and related fields. We discuss foundational user research methods (e.g., needfinding) that are presently underutilized in computer vision and graphics research, but can provide valuable guidance for research projects.  We provide further pointers to the literature for readers interested in exploring other UXR methodologies. Finally, we describe broader open issues and recommendations for the research community. We encourage authors and reviewers alike to recognize that not every research contribution requires a user study, and that having no study at all is better than having a carelessly conducted one. 

\section{Introduction}

Most research in computer graphics and image synthesis produces outputs for human consumption. In many cases, these algorithms operate largely automatically; in other cases, interactive tools allow professionals or everyday users to author or edit images, video, textures, geometry, or animation. For example, photo manipulation algorithms allow artists and casual photographers to modify images for expression and visual communication; geometry synthesis algorithms allow artists to create geometry for video games and movies, to facilitate architectural and industrial design; material models can then be used to texture the geometries; image restoration algorithms, such as super-resolution and colorization, aim to produce visually plausible and appealing images. Likewise, many synthesis algorithms published in computer vision are also designed for human consumption, including \zoyagain{generative AI}, image enhancement, image stylization, neural rendering, and 3D capture of faces and bodies. \zoya{\textbf{When the tools or outputs are meant for user consumption, at what point in the project should users be brought in to evaluate them, and how can the results of user studies further benefit the research?}}
%what is the right way to evaluate them, and how should the evaluations be reported in a paper? 

We have recently seen a proliferation of research papers \zoya{in computer vision and graphics venues reporting} ``user studies'' in which crowdworkers rate algorithm outputs,  \zoya{executed at the end of the project timeline, as an afterthought or in response to the review process. 
On the one hand, when conducted hastily and without sufficient attention to the study design choices, replicability of the published study results can suffer. \zoyagain{We encourage authors and paper reviewers alike to evaluate whether and when a user study is necessary, and to avoid asking for---or running---perfunctory studies that do not affect the paper's conclusions or project directions.}
On the other hand, the true benefit of user studies lies in having them shape the evolution and strategy of a project, or as is the case with foundational research, even the initial project direction. When conducted at the very end, researchers leave no space or time for the results of the user studies to lead to meaningful project improvements or iterations.}
%We argue for elevating the status of user studies in computer vision and graphics paper, in recognition of their ability to .}
%This can result in published studies that are not \textit{replicable}, meaning that another researcher who reran the study with new users would come to different conclusions. 
%Performing good user research is hard; it is better not to do it at all than to publish misleading results. 
%It is crucial for authors and for paper reviewers to understand the limitations of user studies, to evaluate when a user study is necessary in the first place, and to know how to design and evaluate user studies. 

\zoyagain{These considerations are particularly timely with regards to the recent explosion of generative AI technologies. The gap from research iterations to consumer-facing products has shrunk, and users are increasingly being put in front of powerful image and text generation technologies with enormous ethical, legal, and societal implications. In these cases, the types of computational benchmarks common to other facets of vision and graphics research are less relevant, and instead, the focus turns to user behavior, reactions, and interactions with the technology. Here the opportunities for user research are to assess user needs and to forecast user behavior and reception early on and regularly during the model development lifecycle. }

Assuming that researchers want their algorithms to be used in the real world, developing useful tools often requires talking to real users. However, getting meaningful feedback is very difficult and may require specialized expertise.
This discipline of understanding the user, their needs, and feedback is called \textit{user research}, \zoyagain{and was born out of the intersection of psychology and human-computer interaction, pioneered by electrical engineer and psychologist Don Norman}. Many technology companies employ user experience researchers, or UXRs (including coauthors on this paper). % ~\citep{boyd2020}
 While we urge researchers to collaborate with experts---such as UXRs, HCI researchers, or human perception scientists---this is not always possible. \zoyagain{Further, some models and applications (e.g., generative AI) may require testing on a larger user base than would be tractable for qualitative methods. For these reasons,} \textbf{this paper offers a guide and introduction to user research methodologies relevant for graphics and vision researchers}.
For further reading, we provide pointers to key resources on user research, and the terminology for talking about user research that can help navigate those resources.
This paper draws on our own academic and industrial experience with user research, within computer graphics, vision and other areas. \zoya{In providing this background, we hope to expand vision and graphics researchers' repertoire of user study methodologies for gaining different types of insights throughout the project lifecycle.}

We categorize user research methods into three buckets: 
\textit{Output Evaluation} (Section \ref{sec:outputeval}),
used to evaluate the outputs of an algorithm or compare outputs between algorithms;
\textit{Interface Evaluation} (Section \ref{sec:interfaceeval}), used to evaluate how an interactive tool can support or augment a user’s typical workflow or otherwise facilitate task completion; and lastly, 
\textit{Foundational Research} (Section \ref{sec:foundational}), 
performed before any tool or algorithm has been built, to help guide design and development to meet real user needs. This last type of user research  is rare in vision and graphics research, but more common in HCI and corporate product development. We describe techniques for designing effective evaluations, getting more information from studies, and avoiding common pitfalls that may invalidate results or hinder replicability.

%All too often, user studies are treated as an afterthought in graphics and vision research. 
\zoya{In this paper, our goal is to elevate the role of user studies in graphics and vision research.} We argue that they should be treated with the same care and rigor expected of other parts of the research, \zoya{and in doing so, can directly shape the project direction.} We close with a maxim to keep in mind: \textbf{bad user research leads to bad outcomes}, and we discuss ways that flawed user studies can mislead or misguide research and product development.
We  hope this paper will help researchers perform better user studies, leading to useful evaluations and new insights that can inform and inspire their research.
%On the other hand, not all research warrants quantitative user evaluation, which can hinder research progress \citep{greenberg2008usability}.}
%But they should be used in moderation: \textbf{an over-emphasis on quantitative user evaluation can hinder research progress} \citep{greenberg2008usability}. 
%We recommend that paper reviewers restrain themselves from expecting user studies in papers, except with good reason. 

% ----------------------------------------------------
\section{The start of user studies in graphics and vision}

Computer graphics research has long developed algorithms and interfaces for creating and manipulating images, videos, textures, geometry, and animation. Computer vision venues now also actively publish papers on these topics. 
While the graphics community has long discussed the need for rigorous evaluation for more aesthetic and output-focused work \citep{AgrawalaPrinciples,HertzmannScienceOfArt}, 
evaluations of aesthetic and interactive techniques appeared only rarely in the first four decades of SIGGRAPH’s history.  For example, in 2008, \aaron{just before the introduction of crowdsourcing,} only 3 of the 103 papers published at SIGGRAPH 2008 performed studies with human evaluation, and none of these papers reported user evaluations of interfaces, or any user feedback.  
Historically, only a few types of graphics papers leveraged user research, including perceptual studies for rendering \citep{renderMeReal,vangorp,Cole:2009:HWD,serrano} \zoya{and animation \citep{cloneAttack,osullivan2001,osullivan2003}, as well as}, occasionally, for evaluating interactive systems \citep{shugrina2017playful,talton}. 

Beginning in 2008, 
\textbf{crowdsourcing drove the graphics and vision fields toward widespread use of ``user studies''.}
Papers in the HCI literature showed that Amazon Mechanical Turk (MTurk) could be used for user evaluations~\citep{kittur2008crowdsourcing} and perceptual studies~\citep{heer2010crowdsourcing}.   This work inspired new research using MTurk to evaluate algorithm outputs, notably \citet{o2011color}. 
Likewise, computer vision papers in 2008 showed that human computation could be used for data labeling~\citep{sorokin2008utility,spain2008some,vijayanarasimhan2008multi,russell2008labelme}, a purpose for which it became widespread ~\citep{kovashka2016crowdsourcing,deng2009imagenet}; it is also used for evaluating predicted labels~\citep{berg2009finding,cho2010content,li2010building,parikh2010role}.
Crowdwork became widespread in other fields; for example, a highly-cited psychology paper by \citet{buhrmester2016amazon} pointed out that MTurk could supplant the usual practice in psychology of using one's own undergraduate students as the data source, a practice that may have produced many misleading and biased results~\citep{jones2010weird}.
In addition to crowdsourcing platforms, \zoya{it is not uncommon to see the use of convenience sampling or friend-sourcing (involving friends or labmates) in the evaluations of vision and graphics algorithm outputs.}
%computer graphics and vision researchers also sometimes evaluate results much more informally, e.g., asking labmates to rate results in the final hours before a paper deadline.

Crowdsourcing  made it easy to quickly and efficiently gather large numbers of human-provided labels and evaluations that would have otherwise been very difficult to acquire. \zoyagain{While providing a way to collect anonymous user feedback at scale, it has frequently been done without the careful attention to study design and methodology that is practiced in the user experience and perception communities (e.g., properly balanced conditions, statistical measurements of confidence and effect size, careful consideration of confounding factors and sources of bias, etc.)}. %This has contributed to a proliferation of \zoya{studies without careful attention to design considerations including sources of bias, confounding factors, properly shuffled conditions, etc.} 
%informally-conducted studies with little thought paid to whether the results are genuinely meaningful. 
This lack of rigor contrasts with the careful attention paid to benchmark datasets and other types of quantitative evaluation \zoya{which are often prioritized in vision and graphics papers.
\aaron{Performing effective studies online requires careful thought and effort \cite{Cuskley}.}
The true potential of user studies to shape tool development and project direction is underutilized}. Moreover, the field has yet to make use of the broader range of user research techniques available, which could \zoya{offer insights throughout the project lifecycle.}
%expose new research directions.

\zoyagain{In the next sections, we categorize user study designs into those most applicable to graphics and vision researchers for evaluating outputs, evaluating interfaces, and ideating project directions in the first place. Where applicable, we include examples of papers from the literature that have used the different types of study designs discussed.}

% ---------------------------------------------------------

\section{User research methods for graphics and vision}\label{sec:methods}

In computer graphics and vision research, user studies are commonly used for evaluating one of two contributions: the final output of a model or algorithm, which we refer to as \textit{output evaluation}, or the user interface itself, which we refer to as \textit{interface evaluation}. For instance, the typical survey deployed on Amazon's Mechanical Turk (MTurk) or a similar crowdsourcing platform that asks participants to rate which of a set of outputs (images, videos, textures, etc.) are ``better,'' is an example of output evaluation (see Section~\ref{ssec:outputrecs} for recommendations for running these types of studies).

\begin{figure}
  \centering
  \includegraphics[width=1\linewidth]{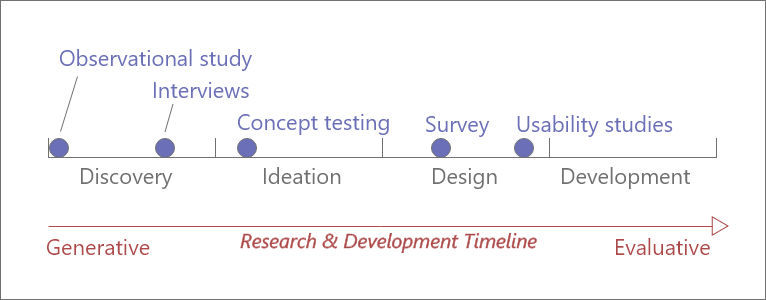}
  \caption{Overview of common methodologies from user experience research relevant to vision and graphics researchers, spanning the research \& development timeline. At the beginning of this timeline are generative methods like observational studies and interviews that help to define the problem and guide the research. At the other end of this timeline are evaluative methods like surveys and usability studies for assessing the final results.}
  \label{RDtimeline}
\end{figure}

One way to think about user study methodologies is to consider where they occur on the research \& development timeline (Figure~\ref{RDtimeline}). This is common practice in corporate product development, but applies well to research. For instance, surveys represent just one possible study design, and together with other output and interface evaluations, occur near the end of the research \& development timeline. \textit{Foundational research} (including methods like needfinding interviews), on the other hand, occurs before any tool development takes place and is used to define the problem statement and guide development. While rarer in graphics and vision papers, we advocate its wider use, as it can help lead toward more impactful projects. The rest of this section is devoted to describing and providing recommendations for output evaluation (\ref{sec:outputeval}), interface evaluation (\ref{sec:interfaceeval}), and foundational research (\ref{sec:foundational}), but we first introduce some terminology.

\textit{User experience research (UXR)} categorizes methodologies using the following axes: 
qualitative/quantitative, attitudinal/behavioral, and generative/evaluative. \textit{Qualitative} research involves collecting non-numerical data to understand behaviors or opinions (“how or why do people do something?”). Conversely, \textit{quantitative} research involves collecting and analysing numerical data (“what percentage of people in a population do something?”). 
\textit{Attitudinal} approaches focus on what users say, whereas \textit{behavioral} approaches focus on what users do. Generative approaches collect information to help form the foundation of the research and development process. Questions such as, “whom are we building this tool for?”, and “what user problems are we solving?” are commonly addressed with \textit{generative} approaches. In contrast, \textit{evaluative} approaches collect information to refine an existing concept or tool. \zoyagain{Discussing study methodologies in terms of these axes makes explicit what a given user study is measuring (e.g., user attitudes or user behaviors?) and for what purpose (e.g., to guide the development of a tool or to evaluate its usability once built?).}
%\todo{what does "brings to the forefront" mean here? what is this trying to say?}

\textbf{What do you hope to learn?}
\zoya{In deciding what types of user studies to run for a research project, the critical questions are:}
%When colleagues come to us for advice on user studies, our first questions are: 
what are your goals, what do you hope to learn, and what data do you hope to gather?
%Answers to these questions can help identify the most suitable study methodologies. 
If the goal is to learn from existing user workflows, then a \emph{generative} approach, such as an \emph{observational study} or \emph{interview} is appropriate. If the goal is to report which algorithm is better, a \emph{quantitative survey}, performed in the \emph{evaluative} phase of the research \& development cycle, may suffice. But for additional insights about \emph{why} users made the selections they did, trade-offs between algorithms, or failure modes of an algorithm, \emph{qualitative} approaches like \emph{interviews} or \emph{focus groups} should be considered. User research can be carried out throughout the duration of research \& development, to test evolving assumptions about user workflows and continue to iterate on tool design. Doing so can also help reduce surprises during final evaluation.

\textbf{Should you perform a study at all?} 
Before embarking on user studies for a paper, we recommend thinking through the study and whether it will provide genuinely useful information. For example, if the study aims to evaluate image outputs, will the study tell you something you cannot tell from looking at the images yourself? If the study will not be performed carefully and rigorously, will the results be meaningful and replicable? If crowdworkers will be used, will their responses be representative of the target users of the tool/algorithm?
Does the technique even merit evaluation, e.g., for early-stage experimental designs, where a meaningful baseline comparison might not possible~\citep{greenberg2008usability}? Likewise, paper reviewers should ask themselves similar questions before making demands on the authors. We discuss these considerations in more depth in Section~\ref{sec:limitations}, where we argue that it is better to perform no user study at all, than to perform a poorly-designed or perfunctory study.

\subsection{Output Evaluation}\label{sec:outputeval}

We use \textit{output evaluation} to refer to evaluations of the final products of algorithms such as synthesized or modified images, textures, animations, or videos, e.g., algorithms designed for inpainting, tone adjustment, image stylization, or texture synthesis. \zoyagain{Examples of output evaluations of generative AI models include~\citep{denton2015deep,zhou2019hype}. Both leverage psychophysics-inspired experiments that display a sequence of images to participants and have them evaluate whether each image is real/fake, while varying image presentation times to compare output quality across models (i.e., an image detected as fake in a shorter presentation time represents a poorer quality model result). Note that this study design does not however evaluate \textit{why} (what aspect of) an output is perceived as fake.}

Output evaluation can be carried out as an interview, a questionnaire, a survey, or an observational study, although surveys are most common in computer vision and graphics papers. In UXR terminology, a \textit{survey} involves
asking study participants to rate or compare alternatives, in order to generate metrics about user ratings.
It is a \textit{quantitative}, \textit{attitudinal}, and \textit{evaluative} study design. 

Properly designed surveys improve the likelihood that the study is replicable.
Beyond the high-level recommendations that we provide below, good practices for survey design can be found in~\citet{rea2014designing} and~\citet{kuniavsky2003observing}. However, there are also whole courses on survey design and psychometrics that UX researchers often take as part of formal training or graduate degree programs, so we recommend consulting with an expert if possible.

\subsubsection{Study Recommendations}\label{ssec:outputrecs}

\aaron{We now describe several interface designs for rating images, several of which are summarized in Figure \ref{fig:rating_uis}.}
Consider the case of two image \zoyagain{generation} algorithms, algorithm A and algorithm B, that need to be compared. This requires a selection of images on which to compare these algorithms, and an interface that will present images to participants, asking them to evaluate \zoyagain{the results, for instance by clicking on the image of the pair they prefer (Fig.~\ref{fig:rating_uis}d)}. Participants will be asked to make some number of judgements in sequence. This type of evaluation study is most typically run by graphics and vision researchers on a crowdsourcing platform, and the results are tallied to report what proportion of participants preferred algorithm A’s outputs. This common design is also referred to as: two-alternative forced choice (2AFC), A/B testing~\citep{kohavi2009controlled}, and randomized controlled trials in science, so similar study guidelines apply.

\newcommand{\uifigscale}{0.25}

\begin{figure}
    \centering
    \includegraphics[scale=\uifigscale]{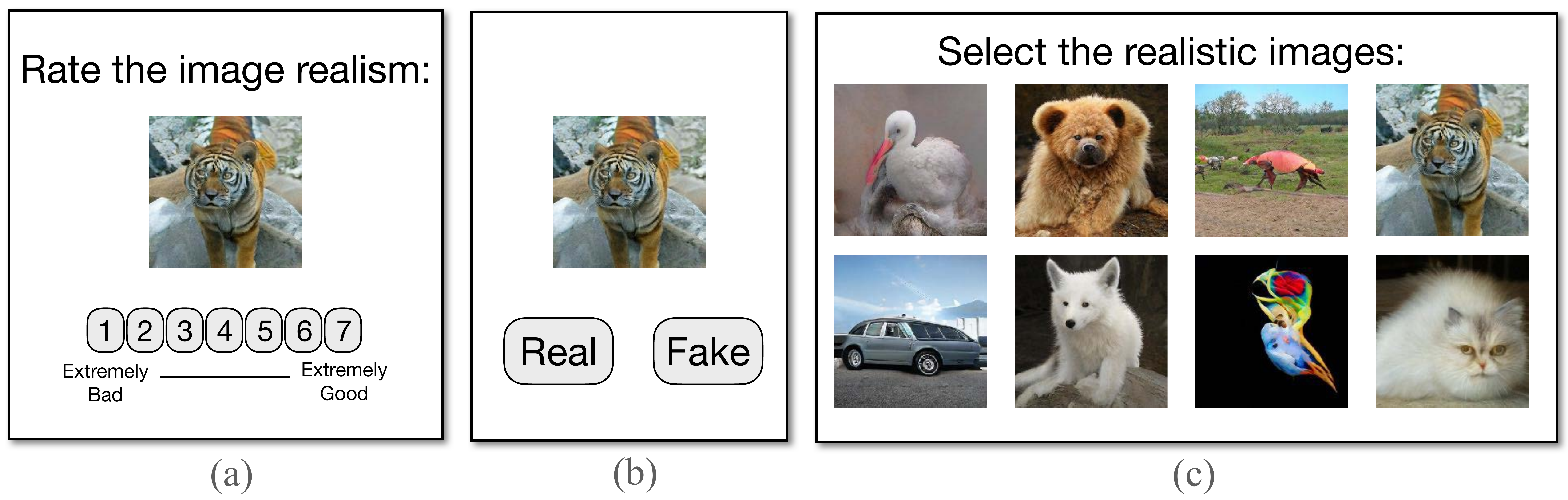}
    \includegraphics[scale=\uifigscale]{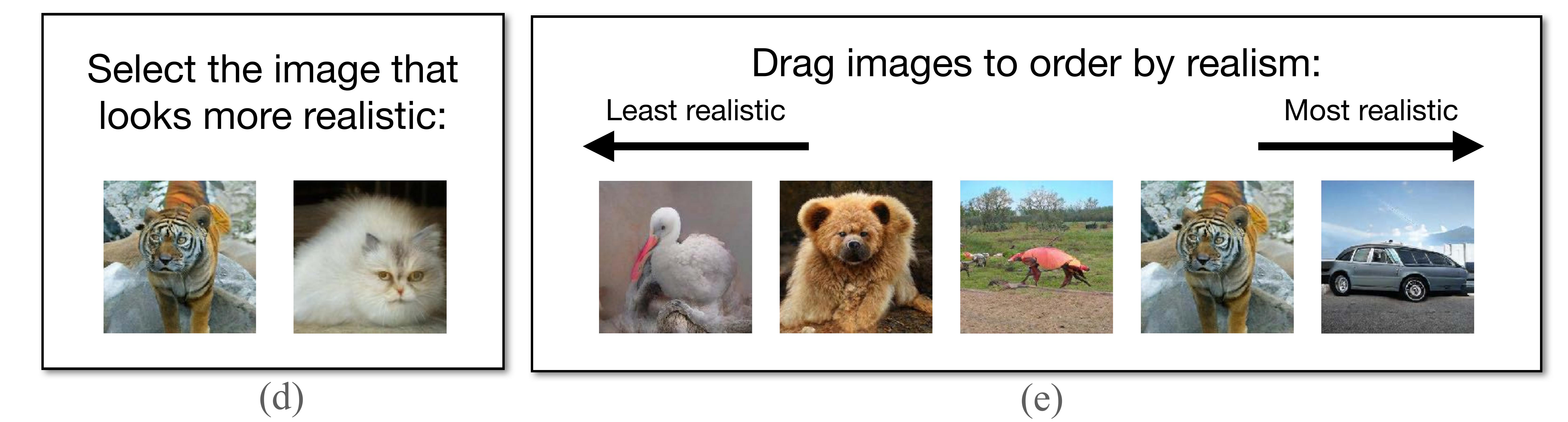}
    \caption{\zoyagain{Sample user interface designs for output evaluations, ranging from single-image assessments and ratings, to A/B comparisons, and multi-image evaluations.}}
    \label{fig:rating_uis}
\end{figure}

When a larger number of outputs or algorithms need to be compared, a pair can be sampled on each trial \zoyagain{(e.g.,~\citep{wang2019planit,ritchie2019fast,lu2012helpinghand})}, or participants may be presented with three or more options simultaneously and asked to order or rank them \zoyagain{(Fig.~\ref{fig:rating_uis}e)}. The average rank assigned to images from a particular algorithm can be used as a summary score (e.g.,~\citep{fosco2020predicting}). \zoyagain{Another option is showing multiple outputs at once, from any number of algorithms, asking participants to select all the images that match a particular criterion of quality (Fig.~\ref{fig:rating_uis}c, e.g.,~\citep{WangIndeterminacy}). During analysis, the proportion of images chosen from each algorithm can be used to compare the algorithms (e.g., \citep{li2018closed}).}
Alternative designs include showing each of the options one by one or all at once, asking participants to rate each option on a Likert scale (Fig.~\ref{fig:rating_uis}a, e.g.~\citet{rosala2020}), \zoyagain{or to make a simple judgement for each image (e.g., real/fake like in Fig.~\ref{fig:rating_uis}b as in \citep{denton2015deep,zhou2019hype})}. Ratings can be averaged across participants and used to rank the algorithms \zoyagain{(e.g.,~\citep{chang2018pairedcyclegan})}.
%There are many ways to perform such a study badly, biasing the results. 
Below we describe techniques for improving the quality of these studies, increasing the likelihood they give useful, replicable results, and avoiding common pitfalls.

\textbf{Select a fair sample of outputs; the more, the better.} To fairly compare two algorithms, example outputs should be randomly selected, rather than hand-picked. The exception is if different edge cases need to be explicitly represented (e.g., an equal distribution of natural scenes, cluttered scenes, portraits, etc.). The image selection criteria should be clearly reported for replicability. %The more example outputs evaluated, the better. 
A \emph{within-subjects} study design would be one in which each participant sees all conditions\zoyagain{, and requires a different analysis than a \emph{between-subjects} design that shows different conditions to different participants (e.g.,~\cite{zhou2019hype,lu2012helpinghand,lu2013realbrush}). }
Across evaluators, it is important to shuffle the sequence of presented images to avoid any ordering effects (e.g.,~learning/ramp-up, context effects, fatigue, etc.,~\citet{elmes1999book}). 

\textbf{Anonymize the study.} Whenever possible, study participants should not know which of the outputs is the authors’ method. Otherwise, participants may respond with the answers they think the researchers want, consciously or not, which is known as the “good subject effect” (e.g.,~\citet{nichols2008good}). Do not label outputs with names like “our method” or “existing method”. Participants can be biased by power dynamics (i.e., the researcher holds power by running the study), researchers using language to prime participants (e.g., “how much do you like this tool I built yesterday?”), and researchers and participants’ relationship (e.g., if both work in the same lab or company).

\textbf{Consider task presentation.} It is important to randomize the presentation order of the algorithms: if, say, in a side-by-side comparison the baseline is always shown on the left as image A and the new algorithm as image B, participants may infer that image B is the new method and prefer it; or they may even get in the habit of just clicking B rather than comparing each image pair separately. Other presentation aspects such as the size of the images on the screen, their distance to each other, etc. may influence participant responses. Piloting the study with a few different settings may help spot these potential confounds early\zoyagain{, and provide an estimate of participants' answer variability that is due to study design, independent of algorithm performance}. 

\textbf{Add checks throughout.} Remote crowdworkers may click randomly to speed through a trial and get paid, rather than answering questions in good faith. \aaron{Moreover, recent evidence indicates a substantial decrease in MTurk data quality in the past few years, likely due an increase of bad actors and bots \cite{Chmielewski}.} For these reasons, it is important to build in checks throughout the study design (e.g.,~\citet{sauro2010}). Checks (also called \textit{sentinels} or \textit{validation trials}) are trials for which there is an objectively correct answer that can be used to filter out poor-quality data \zoyagain{(example validation conditions described in~\citet{wang2019planit,lu2012helpinghand,lu2013realbrush})}. For instance, in the case of pairs of \zoyagain{generated} images, one image of the pair can be an intentionally and objectively poor quality result. During analysis, data from participants that failed some preset number of the checks can be discarded, assumed to be generated by participants that were inattentive or inconsistent. 
These checks should be randomly inserted in the study, and should appear the same as other trials; otherwise, participants may figure out which trials are the checks. 
Any data exclusion criteria must be reported for replicability. 
For studies where answers are subjective, such as aesthetic preferences, another option is to duplicate some trials and verify that the participant's answers are self-consistent. However, there should be enough checks to rule out the possibility of good-faith participants answering randomly on examples that are genuinely difficult to judge.

\textbf{Choose wording carefully.} The way the task or question is phrased can also significantly affect the results \citep{epstein2020gets}. Wording should reflect the high-level goals, 
e.g., “which image contains fewer artifacts?” instead of “which image contains fewer color defects in the facial region?” Conversely,  imprecise task wording leaves too much to interpretation, e.g., “which image is better?” may be understood as “which is more aesthetically-pleasing?” where the intention might have been to evaluate “which is more realistic?” Pilot studies with different wordings can help reveal potential biases due to question wording.

\textbf{Sample participants representatively.} It is important to recruit participants representative of the target user profiles who will ultimately use and/or be impacted by your technology. For instance, if building software for animators, recruit animators to provide feedback. It is also important to recruit inclusively, including, for example, people with disabilities. A common approach of convenience sampling or friend sourcing, whereby a researcher leverages an easily-accessible population, often on a volunteer basis, can lead to biased results that won’t replicate.

\textbf{Augment ratings with qualitative feedback.}
Running a survey to get quantitative scores does not provide any information about \textit{why} participants gave certain scores. 
We recommend follow-up interviews or questionnaires to understand \textit{why} participants gave the responses they did.  These may help debug the study, but they may also reveal unexpected facets of the output that influenced users’ ratings. Was the participant ``very unsatified'' with the output because it was unrealistic, not aesthetic, biased, or for some other reason?
\zoya{Without this information, the researcher may end up optimizing features of the algorithm that don't address, or may even amplify, the underlying user problems \zoyagain{(e.g., focusing on improving realism at the expense of less variable or more biased results).}}
%work on refining the algorithm to be more realistic, instead of addressing the underlying user problem.  

\subsection{Interface Evaluation}\label{sec:interfaceeval}

Examples of interfaces that appear in computer graphics papers include creativity support tools that help users create graphic designs (e.g.,~banners, logos, illustrations) or interfaces that help users sculpt 3D meshes. \zoyagain{For generative AI tools entering mainstream use, the user interface becomes key to how users interact with the model and specify their needs, expectations, and desired outputs. }
When the interface itself is a research contribution instead of, or in addition to, the outputs that it generates, then an \textit{interface evaluation} should be performed. In UXR terms, this is an \textit{evaluative} approach which can yield both \textit{behavioral} and \textit{attitudinal} data. Interface evaluation can be used to collect \textit{qualitative} feedback and \textit{quantitative} data (e.g., number \zoyagain{and location} of clicks, time-to-complete, eye gaze, etc.). \rev{Ideally, both \textit{qualitative} and \textit{quantitative} approaches are combined (i.e., \textit{mixed methods}) to produce a robust understanding of how the interface is used.}

In a typical study, participants may be recruited to test the functionalities of the tool while generating sample outputs. Methodologically, this may take the form of a usability study or a concept test. A \textit{concept test} is used to examine initial interface design approaches; it typically occurs earlier in the research \& development timeline (Figure~\ref{RDtimeline}). Concept testing involves participants walking through an early interface design concept: not every button needs to be perfected nor every path through the experience mapped out. Instead, a concept provides the general gist of an experience, which can be shared with users to glean both \textit{behavioral} data (collected through observation of participant interaction with the prototype) and \textit{attitudinal} data (collected through questioning users about their experience with the prototype). A \textit{usability study}, on the other hand, is used to glean concrete, low-level insights into how well users are (or are not) able to use the tool. A usability study is typically performed after an interface has been designed and implemented, at the end of the research \& development timeline. Some quantitative metrics might be captured, like the time or number of clicks/selections required to complete a sample task with this tool \zoyagain{(e.g.,~\citet{jones2021shapemod})}. Qualitative user feedback might also be recorded. Occasionally, a comparison is made to a related tool or an existing workflow in absence of an alternative tool. Interface evaluation may reveal issues with users' interaction patterns, requiring adjustments to the interface design.
This evaluation is most useful if the researchers can act on the insights gathered to improve the tool.

Interface evaluations published in the graphics literature almost exclusively report quantitative measurements, e.g.,~\citep{augmentedAirbrush,humanintheloop,probabilisticReasoning,taltonKoltun,chaudhuri2013attribit}, although occasionally subjective impressions are also gathered from participants or evaluators, e.g., \citep{philbrick2022primitive,jones2021shapemod}.
In the HCI literature, \zoya{it is more common to see graphics interfaces evaluated with qualitative techniques \citep{Hartmann2020VDE}}. \zoyagain{Recent contributions on generative AI include interfaces for qualitative evaluation~\cite{assogba2023large}.} \rev{Both quantitative and qualitative techniques are beneficial for evaluating interfaces, and we recommend gathering both qualitative and quantitative data when possible.}

\zoyagain{Occasionally, an interface is used to evaluate the effectiveness of an underlying model or algorithm, rather than the interface itself. The goal in these cases may be to compare different approaches using the same interface, and report metrics like sample task completion time, diversity of generated outputs, etc. (e.g.~\citet{probabilisticReasoning}). Most of the same study recommendations apply in these cases, with the exception that the learnings from the study are not used to improve the interface.}

\subsubsection{Study Recommendations}

Consider a tool to help novice users ideate on icon designs. To evaluate the success of the tool, it is important to understand novice users' use of the tool, whether it is indeed usable by novices, and whether it helps them to create icon designs.

\textbf{Recruit study participants from the target populations.} Participants from all the target user groups (e.g., novice, intermediate, advanced users) should be invited. A \textit{screener survey} that queries potential participants’ traits and demographics may be useful to determine whether they fit within a target user group. Participants should be recruited in an ecologically valid manner (i.e., in a way that is reflective of real-world scenarios). This includes finding people who would otherwise naturally come across the tool if it existed, rather than relying on friends, family, or co-workers, who may be biased. Each participant should be scheduled for a one-on-one session in which they will be given a task to accomplish with a working prototype of the system. 

\textbf{Understand the participant’s context.} At the start of the session, participants should be asked questions about the context in which they might use such a system, understanding their goals, motivations, processes, and scenarios in which the system may be used eventually. Ideally, the task and context will be as realistic as possible.

 \textbf{Capture first impressions.} After the broad introductory questions in an interview, participants can be provided with the task and prototype. Gathering first impressions on the task/how to navigate the prototype, after providing minimal to no guidance, can provide valuable insights about users’ natural understanding of the tool being studied. This approach may reveal participant expectations, mental models, and potentially, new UX opportunities. 

 \textbf{Give participants control.} Placing participants in full control of the prototype, without providing task guidance or next steps, can reveal how users naturally progress through the task using the interface. This approach may address questions like: do participants get stuck? Do they know where to go next or how to operate the tool settings? The goal is to learn how participants would approach the task using your tool if you were not available to guide them. 

\textbf{Provide guidance if appropriate.} Sometimes the research goal requires guiding participants to a specific part of a design or prototype where feedback is most desired. For example, if working with a low fidelity prototype, it may not be informative to have participants freely explore the interface.
It might be reasonable to assume that outside a research context, help documentation or an onboarding tutorial may be available to a user before reaching the main interface. In these cases, comparable guidance can be provided. 

\textbf{Test a workflow.} If the research objective involves testing a hypothesized workflow, the participant can be guided through the workflow, carefully assessing user expectations and needs at each step. For example, if a participant gets stuck on a part of the interface, before supplying guidance, they can be asked ``what do you expect to do next?'' If they have reached an impasse and have no further feedback, they can be given hints about what to do next, and asked to comment on the action (e.g., was it easy or difficult, and why?). 

\textbf{Ask \zoya{lots of} questions.} Instead of directing or guiding the process, the researcher can ask questions---e.g., why the participant chose a certain tool, what they’d expect to do next, how they expect to do that, where they might look to find what they need, if the prototype is working as expected, etc. Here, it will be especially important to avoid leading questions, as they may guide the participant to make unrealistic choices. For instance, a leading question may direct the participant to a menu that would be otherwise hidden, or it may cause the participant to take a step that they would otherwise omit. This may result in overly optimistic results. After the participant completes the task, it is important to ask for feedback on the overall experience, which may be surprisingly different from the feedback provided during the task itself. It’s helpful to ask what the participant would change about the experience, as well as if, when, and how they would use the system.  

\textbf{Collect data for data-based claims.} There is a plethora of data that may be collected throughout the interface evaluation study.
Participants' attitudes towards using the prototype for the given task can be collected as qualitative data (constructive feedback). How the participant interacts with the prototype can also serve as qualitative behavioral data, e.g., where do they indicate confusion by pausing or asking questions? Quantitative behavioral data can be captured by measuring participants’ number of clicks, time-to-complete, eye gaze, etc. This data can be aggregated across participants to deduce trends; additionally, similar measurements can be collected from participants’ interactions with existing or competitor systems to drive statistical comparisons.
Aside from fueling data-based claims about the efficacy of the designed system, interface evaluations also provide concrete feedback to improve the user experience of the system; in the case of a concept test, this feedback is harvested early in the design process to ensure that the ultimate interface is usable and effective.

\subsection{Foundational Research}\label{sec:foundational}
Compared to evaluations of interfaces and outputs, \textit{foundational research} occurs prior to the start of development to learn more about a problem space, and help identify potential audiences and use cases. These insights provide direction to design, research, and development. 

In UXR, the most well-known type of foundational research is \textit{needfinding}: the process of engaging with potential users to uncover opportunities, through interviews and by observing their behaviors in context. It is a \textit{qualitative}, \textit{generative} approach that can yield both \textit{behavioral} and \textit{attitudinal} data. Needfinding typically occurs at the outset of a project, informing what the team ultimately builds, sometimes in ways that would never have occurred to the researchers or developers otherwise.
Needfinding plays a crucial role in the iterative design process for commercial product development, which has been shown to lead to more successful products \citep{Forrester,schaffhausen2015large,rosenthal2006ethnographies,von1986lead}. We believe these benefits will apply to academic graphics and vision research that aims to benefit real users. Needfinding is widely used in the HCI literature. 
There is also value in exploratory research based on the researcher’s own intuitions, but with caution to be wary of the false-consensus effect~\citep{nielsen2017,ross1977false}, i.e., overly generalizing from one's own experience to others. 

Indeed, we have often witnessed computer graphics or vision researchers attempting to get their research adopted by industry practitioners, only to find that the research does not address the target users' needs. Researchers who do not perform needfinding at the outset may be surprised to find that users have no need for or interest in the tool they have spent months or years developing. Such tools may perform poorly in evaluation studies, as users may find that the technology produces unhelpful, irrelevant, or unexpected results. Needfinding can help build technologies that will perform well when evaluated by real users, and may be more likely to find a home in commercial tools. 

Foundational research can also gather \textit{generative} insights such as design principles and expert workflows, from experts or ``lead'' users in a field~\citep{goodman2012observing}. These insights can be used for developing tools for either experts or novices. For experts, foundational research can identify stages of existing workflows that can be improved with new algorithms or interactions.
For novices, design principles gleaned from experts can be embedded into software tools or can provide automated guidance. An example is the survey by \citet{AgrawalaPrinciples} of effective visualization tools designed by extracting principles from expert-drawn illustrations.

\subsubsection{Examples}

Since needfinding is less common in the graphics and vision literature, we describe two research projects that benefited from it, based on interviews with the paper authors. In each case, needfinding provided important insights that guided the direction of the research; the authors believe that needfinding made a significant difference in their ability to produce useful systems. %, rather than achieving a “made-up” goal.
Our first example is in digital painting. Motivated by the limitations of conventional color pickers for digital painting, \citet{shugrina2017playful} sought to create a better color picker and mixer. The researchers conducted a needfinding study to examine how artists used paint palettes in their daily work, to discover the properties that a tool should have to support these workflows. Based on these studies, the authors designed their Playful Palette tool to combine the physicality of oil and watercolor palettes with digital affordances.
The authors then evaluated their tool with another group of artists, demonstrating that it was useful and effective for supporting their creative tasks. 
They told us that ``with a questionnaire, one can struggle to get insightful answers from participants, but observing them do something can uncover their processes, including possible struggles and inefficiencies". By starting with an understanding of users' actual needs, the authors were able to build a system that met those needs and achieved positive results in the evaluation.

Our second example comes from \citet{zhao2020iconate}, who built a tool for novices to partially automate icon design. The authors started by studying professional icon designers to gain insights about what aspects of their workflows could be automated. They collected attitudinal data through one-on-one interviews and behavioral data through user observation sessions. 
These studies revealed that professional designers often collect references from various websites, and from these reference images recombine visual features into the final icon design, later refined using vector editing operations. 
Informed by these observations, the researchers built the Iconate tool to automatically suggest recombined icons for various user queries. 
Usability tests with novices provided qualitative data that users liked the icons they generated with the help of the tool, and quantitative data supported that they were faster than when using other conventional tools. Evaluative interviews with professionals demonstrated that they too found the tool useful. The difference between the two target user groups was that professionals only wanted the tool for prototyping ideas, but preferred to complete their designs in external vector editing tools, while novices were keen to use the tool for generating the final icon designs. Over the course of the initial and final interviews, the tool both improved to better fit the needs of the target users, and the paper was more strongly motivated with user needs in mind.

\subsubsection{Study Recommendations}

Foundational research should come at the start of a project, and will be most beneficial before any technology has been created. Needfinding, the most common type of foundational research, is composed of four core stages~\citep{patnaik1999needfinding}. In the first stage, \textit{framing and preparation}, one determines fundamentals such as the research goals and the target, and performs a literature review. In the second stage, \textit{observation and recording}, the researcher immerses themselves in the target audience’s context to observe their behavior. The researcher must consider how their presence in the audience's context might affect behavior, and take measures to minimize disruption to individuals' natural behavior. Third, in \textit{asking and recording}, the researcher interviews the target end users in their own environment, directly recording their words. Last, during \textit{interpretation and reframing}, data is framed in terms of the problems the users need to be solved to improve their situation.

\textbf{Identify the target users.} Identifying an audience or target user group for one’s technology at the outset of the project will naturally focus the development and lead to better design. A tool that is targeted too broadly may not end up satisfying the needs of any group. For instance, a graphic design tool built for novice users will likely have a different set of attributes than one for professional graphic designers. Professionals might not want a one-click/one-touch solution, but to have much more control over the creative process, compared to novices who may benefit from fewer options and more automation (e.g.,~\citet{zhao2020iconate}). Similarly, in designing a tool for photographers, there may be important differences in user needs and workflows, for professionals versus amateurs, or portrait versus landscape photographers. Needfinding can help discover these differences and identify the best target user segment for evaluative research. 

\textbf{Recruit users in the wild.}
Once a target user group has been selected for study, participants should be recruited in naturalistic settings, i.e.,
where they naturally congregate. For example, there may be a relevant active online message board or social media community that can be leveraged. Even meeting one member of the community may lead to \textit{snowball sampling}, in which relevant users offer connections to similar individuals in their network. 

\textbf{Observe users in their natural environments.} In an observation study, there is limited interaction with the users; one simply observes users going about their standard processes, while noting any apparent ``pain points'' or ``aha'' moments. These insights help inform the technology that will be built---it should close a gap, address a pain point, or enhance a current process. Prior to observation, consent may need to be collected, considering institutional guidelines. 

\textbf{Seek to understand the users' contexts and needs.}
One-on-one interviews with target users can be conducted in lieu of or in addition to an observation session. In these interviews, researchers should start off as broadly as possible, seeking to understand users and their contexts. Examples include mapping out their current workflows, and probing their motivations and decision making processes. If the goal is to understand a specific workflow, users can be asked to walk through it step-by-step. Screensharing can be used with remote users. It is useful to combine interviews and observational sessions (a methodology called \textit{contextual inquiry}) to best understand target users from both \textit{behavioral} and \textit{attitudinal} perspectives. This may also take the form of an initial interview, followed by an observational session. Asking participants questions about their process as they perform it in real-time is a convenient way to probe certain areas of interest.

\textbf{Minimize bias during your interviewing.} The way in which a researcher structures an interview and asks questions can bias participant responses. First, it is important to make the participant feel as comfortable as possible by clarifying that there are no right or wrong answers. Making space for silence instead of finishing the participant's sentences, and avoiding providing feedback throughout the interview (even saying “I agree”) will help minimize bias in their answers. Leading questions should be avoided (as discussed in Sections~\ref{sec:outputeval}, \ref{sec:interfaceeval}). Other types of interview questions to avoid are future behavior predictions (e.g. “What do you think your process will look like in the future? Would you ever use a tool like this?”); numerous psychology papers (for reviews see~\citet{armor1998situated,dunning2007self}) have taught us that humans can not accurately predict their future behaviors. Double-barrelled questions (e.g., “What do you think about performance, speed, and accuracy? How do you decide which tool to use on a given day?”) should be broken up, to ask one question at a time (the example above could be four separate questions). 

\textbf{Write a discussion guide.} One way to minimize bias is to prepare and review the questions in advance. A \textit{discussion guide} is a set of user-facing questions, distinct from the study’s internal research questions, formulated specifically to learn relevant information from participants. The first questions in the guide should include some “warm-up” questions, not specifically tied to the goals of the research (e.g., ``Where are you calling in from today?'', ``What do you like to do for fun?''), to help participants become more comfortable with the interview. These questions may also provide some insight into the participants' context and what is important to them. The discussion guide questions can then get increasingly narrower in scope, including questions about workflows, pain points, and finally, honing in on participant reactions to the presented solution. 
The discussion guide includes a list of questions to choose from to provide structure, and ensure the research goals are addressed, but is just a guide. We recommend taking a \textit{semi-structured} interview approach, which means the researcher can follow-up on certain answers or ask new questions as they emerge. 

\textbf{Consider your own assumptions and biases.} The researcher’s perspective is influenced by their role in conducting the research, their own experiences, and assumptions~\citep{cohen2002research}.
\textit{Reflexivity} refers to examining one’s own beliefs, judgements, and practices during the research process, and how these factors can impact the research~\citep{finlay1998reflexivity}.  Limitations identified should be noted by the researcher throughout the research process, and considered during data analysis. 
This practice may shape how we approach reporting results, including contextualizing the research through a particular lens, or reporting biases as limitations.

% ---------------------------------------------------------

\section{The broader landscape of user research methodologies}

There are many ways to gain important insights from human participants throughout the research \& development lifecycle, from conceptualizing a tool to iterating on the final designs. Depending on the research goals and the stage of the research, different user research methodologies may be appropriate. In this section, we provide an overview of this broad landscape of methodologies (Figure 2), with pointers to further reading.

The type of data one will need to address their objectives, questions, and decisions can guide the choice of methodologies. For instance, attitudinal data (i.e., what users \textit{think}), can be collected with a survey; while behavioral data (i.e.,~what users \textit{do}) can be recorded during observational sessions. 

Considering the current state of the project will further narrow down the relevant methodologies. Foundational research is useful during early ideation, while output evaluations are conducted with a fully-built model. Generative methods are conducted in the early, discovery stages; they can help define a target user to guide research or product development. Conversely, if the target user has already been established, and the product concept validated, a usability study can evaluate the implementation. Evaluative methods generally apply when assessing functionality, from an early-stage concept to a fully-built product, and are typically used in the later stages of a project.

\subsection{Other study designs}
Table~\ref{tab:definitions} provides a working definition for the user research methods we posit are most applicable to graphics and vision researchers (i.e., in the darker color in Figure 2), and which we already discussed in Section~\ref{sec:methods}. 
In the lighter color are other frequently used UX research methods that are currently underutilized by vision and graphics researchers. 
For instance, \textit{focus groups} allow several participants to be simultaneously present, engaged in group discussion facilitated by a moderator. Focus groups are well-suited if a researcher has already identified core themes they want to further explore with participants. Such themes may emerge from earlier observational studies or one-on-one interviews. \textit{Diary studies} are particularly useful for reviewing a participant's experience over time, and tracking behaviors (e.g., software usage) over a multi-week period. Participants are typically supplied with pre-defined prompts or questions to answer throughout the study. \textit{Participatory design} is a process by which researchers and participants co-create an interface, not to be used as-is, but to reveal hidden user needs and considerations. \textit{Eye tracking} can be used to track and show where a participant is looking, typically to understand how the participants process a given user interface.

\begin{figure}
  \centering
  \includegraphics[width=1\linewidth]{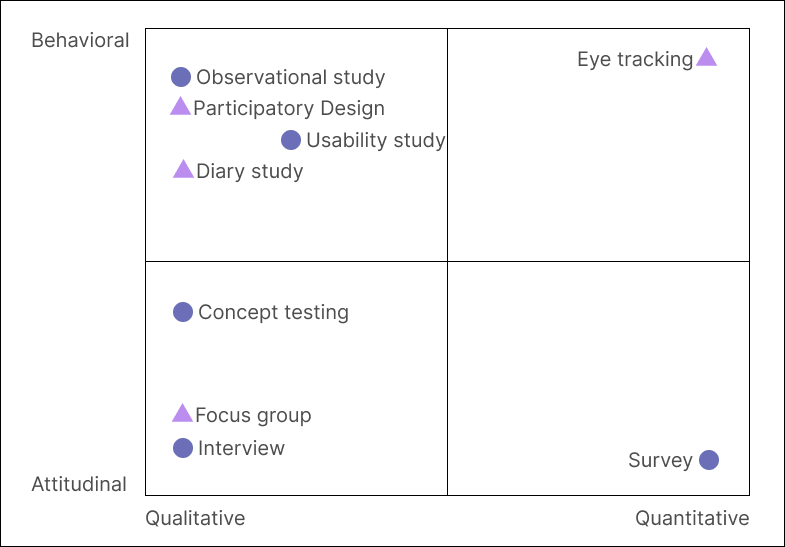}
  \caption{\zoyagain{Dark color circles} are the methods discussed at greater length in this paper, and \zoyagain{lighter color triangles} are other relevant methods that are mentioned in passing. The user research landscape is larger than represented here, with more study designs possible, but these particular examples were curated with the vision and graphics research audience in mind (adapted from~\citep{rohrer2014}).}
  \label{UXRmethods}
\end{figure}

\begin{table}
  \caption{Definitions of common methodologies in user experience research. Further reading on user research methods include~\cite{creswell2017research,goodman2012observing,sauro2016quantifying}. }
  \label{tab:definitions}
  \begin{tabular}{cp{12.5cm}}
   \toprule
    Methodology & Definition\\
    \midrule
    Observational study & Observation of participants in their naturalistic settings, without any researcher intervention or manipulation of variables. Also referred to as “field research”. \textit{Further reading:~\citet{nielsen2002}}\\ \midrule 
    Interview & One-on-one conversation between the participant and researcher, in which the researcher asks qualitative questions to the participant to collect attitudinal data. Could be used as building blocks in other methods (e.g., concept testing). \newline \textit{Further reading:~\citet{arsel2017asking,portigal2013interviewing}}\\ \midrule 
    Concept testing & Using designs, often of low fidelity, to test concepts, conducted early in the research and development process, serving as guidance for broader project direction. Provides information about whether you are solving the problem you hypothesize you are solving. \newline \textit{Further reading:~\citet{sharon2016validating}} \\ \midrule
    Usability study & A method to evaluate the way in which a participant interacts with a user-facing interface; typically, this method examines the ease-of-use, understandability, performance, and user experience of an interface. This method is typically conducted late in the research and development process, after the concept direction has been validated earlier in the process, via concept testing. \newline \textit{Further reading:~\citet{albert2013measuring,rubin2008plan}} \\ \midrule
    Survey & A method to collect self-reported user data at scale, addressing questions like “what percentage of users do x or y?” \newline \textit{Further reading:~\citet{rea2014designing}}\\
    \bottomrule
  \end{tabular}
\end{table}

\subsection{Alternatives to formal user studies}

While the previous section lists formal user research methods, there are ways to capture usability issues \emph{without} running a user study, e.g., with a usability inspection method~\citep{nielsen1994usability}, such as a cognitive walkthrough~\citep{rieman1995usability} or heuristic evaluation~\citep{nielsen1990heuristic}. In a \textit{cognitive walkthrough}, the researcher puts themselves in the shoes of a potential user \zoyagain{(e.g., see Section 5 in~\citep{assogba2023large})}. The researcher attempts to use the tool, and as they proceed, they evaluate the usability of various potential paths through the user experience. In this process, they consider how users will access various features, where and when they will be able to proceed, potential areas of confusion, etc. A \textit{heuristic evaluation}, while similar, is conducted by an expert and relies on pre-defined quantitative metrics. The heuristic evaluator also steps through the user experience, but they rate the experience on a predefined set of established heuristics (e.g.,~\citet{nielsen1990heuristic}) as they proceed. This is particularly useful for highlighting usability issues and areas of improvement in the user journey. Such evaluations can replace a formal user study under limited resources or be used as a precursor to further user testing. Learnings from such evaluations can be used throughout research \& development to refine tools.

\subsection{Combining methodologies}

Synthesizing the results of multiple study methodologies, via a \textit{mixed methods} approach, is likely to provide richer insights. Including both behavioral and attitudinal, as well as qualitative and quantitative data, will provide the most holistic picture of the user.

It is common for researchers to over-rely on surveys as a quick way to get quantitative data. 
There is a widespread tendency to trust quantitative data over qualitative data~\citep{mau2019metric}, which may be exacerbated by an unfamiliarity with qualitative techniques, a point we hope to address with this paper. Simply adding open-ended questions into a quantitative survey does not harness the qualitative data we are referring to; most participants will not spend the time writing out responses to open-ended questions the same way they might extrapolate on their needs in a 1:1 interview.

\zoyagain{Another natural outcome of formative user studies could be the criteria that users may judge a tool or output by. Interviews and focus groups can give researchers insights on features of a tool or output that are most important or relevant to the users and their workflows, which can then be converted into criteria or metrics for evaluation.}

Consider the example of measuring creativity or evaluating creative outputs. A single quantitative measure will not capture the complexity of what it means for an output to be more or less creative. Past research on creativity-related technology has utilized a bevy of disparate evaluation strategies. One oft-cited study~\citep{greenberg2008usability} concluded that more diverse alternatives will afford a more creative outcome, leading many researchers to prioritize quantity as a proxy for creativity. Other researchers took this a step further, arguing that more efficient tools will enable the creation of more design outputs~\citep{davidoff2007rapidly}, thereby fostering creativity. This approach prizes speed over quality by measuring time-to-complete as a proxy for creativity. Yet other researchers have hired \textit{design experts}, a tricky pool to define and recruit, to arbitrarily rate the quality of creative outputs~\citep{merrell2011interactive}. None of these approaches single-handedly capture creativity, necessitating a mixed methods approach \citep{crilly2019methodological}. As an example, \citet*{herman} propose a mixed methods framework for measuring creativity, in which qualitative, quantitative, behavioral, and visual data are triangulated to formulate a user-centered evaluation of creative tools.

\section{Reporting and conducting ethical research}

\rev{The danger of treating user studies as a second-class citizen is that important methodological details may be left out and terminology may be used incorrectly, impeding future reproducibility efforts. Further, providing too few details hinders the ability to evaluate whether the conclusions made based on those user studies are valid and reproducible. In this section we advocate for careful reporting of methods and conclusions, and for ethical procedures to be followed when working with  participants. }

\subsection{Reporting requirements}

In computer science paper writing and reviewing, it is common to pay laser-focused attention to the description and reproducibility of algorithms. On the other hand, user study descriptions are treated as an afterthought. We argue that it’s not enough to perform a study carefully, it must also be reported thoroughly, so that the reader can understand what was done, what the results show, and how they can be reproduced. Authors should also take care in how they present and interpret their own results.

\textbf{Report methodology reproducibly.}  Descriptions of user studies should include enough details for reproducibility, including how participants were recruited, a description of the participants (e.g., occupation, experience level in relevant domain), stimuli used (e.g., interface prototype, number and type of images), procedural details (e.g., interview length, structure, examples of questions asked)~\citep{schoch2020problem}. Since the exact question wording can greatly affect participant responses (Section~\ref{sec:methods}), this information should be made available, for instance in supplementary material and/or in online repositories.  

\textbf{Use precise terminology.} Computer graphics and vision papers typically use imprecise terminology around output evaluations, beginning with the term \textit{user}. 
The term \textit{user study} originated as a description of an evaluation with prospective users of a real system, for example, an evaluation of word processing software with people who might want to write documents. The crowdworkers in computer graphics and vision studies would more accurately be described as \textit{participants}, \textit{raters}, \textit{evaluators}, or \textit{labelers}. Likewise, the term \textit{user study} as a blanket description for all studies lacks precision, leaving the actual type of study opaque. 
Referring to a study’s design by a precise name can facilitate both study planning and reproducibility, by directing researchers to relevant and related work. Table 1 lists a set of methodologies that all fall under the umbrella term \textit{user studies}, and includes references for further reading on best practices. %, since more precise terminology will help readers better understand the nature of the study being performed before reading the details. 
The present blanket use of the term \textit{user study} in graphics and vision papers gives the misleading impression that all user studies are basically the same. 
When possible, we recommend using more specific terminology. For example, for the common evaluation of outputs done with surveys, we recommend “human rating study”. % as a clearer description of the study design.

\textbf{Interpret results carefully.} One danger occurs when authors over-generalize from an experiment with a limited set of stimuli and narrow participant pool to very broad and general conclusions.  
\aaron{For example, one study in a psychology journal \citep{Gangadharbatla}
concluded that ``individuals are unable to accurately identify AI-generated artwork,'' from comparisons of GAN-generated images against a few existing artworks, made by crowdworkers viewing low-resolution, out-of-context imagery.}
%, images of traditional artworks and images created by AI technologies were gathered from the web, and crowdworkers were asked to distinguish which images came from which sources. From the results it was concluded that ``individuals are unable to accurately identify AI-generated artwork,'' a very broad conclusion that does not follow directly from the experiments, which were performed on a few out-.
This paper does not report details about which specific image sets were collected or used, making the claims hard, if not impossible, to verify and reproduce. More worrisome is that the popular press reported these results with the misleading claims that AIs can make art as well as humans.

% ---------------------------------------------------------

\subsection{Conducting research ethically}\label{sec:logistics}

The previous sections outlined high-level principles for user studies. There are also core ethical aspects that should be part of every user study, including obtaining consent, compensating fairly, and conducting safe and inclusive research.

%\textbf{How to plan?} At the outset of the research, write a study plan. This document will contain background (e.g., literature review and/or business context for the research, if applicable) research objectives, questions, assumptions, hypotheses, and methodological information (e.g., participants, stimuli, procedure). In the context of product development, it is also important to list what decisions are to be made with this research, any major milestones, and stakeholders who need to review the study plan. Stakeholders should include anyone who will be affected by the findings of the research. 

%\textbf{How to recruit?} Recruiting the right participants is key to a successful user study.~\citet{ko2015practical} discuss three recruitment challenges in the context of conducting experimental, quantitative research, namely (1) finding a representative sample, (2) convincing enough participants to join your study such that one has adequate statistical power, and (3) accounting for large variations in skill level across participants. For qualitative research, both (1) and (3) are applicable. The authors also provide strong guidelines for recruitment of software professionals that are also applicable to qualitative methods. 

\textbf{How to get consent?} Different institutions have different requirements (e.g., Institutional Review Board approval) for dealing with human participants. It is important to follow this guidance, procuring informed consent if necessary. If an institution does not require an approval process, it is still highly recommended to receive proper consent from participants before they participate in studies. Participants should be informed of the intent of the study, what data will be collected, how it will be used and stored, and any foreseeable harms or risks. It is good practice to also verbally communicate this information to study participants and ask for necessary permissions (e.g., to record, take pictures, etc.). This helps protect participants and make them feel more comfortable. Getting consent is part of a broader practice of conducting ethical research with human participants (e.g.,~\citet{jhangiani2020moral}, pp. 46-66; pp. 59 for more information on informed consent). Further, all \textit{personally identifiable information (PII)} collected from participants should be anonymized (unless the study design requires otherwise) and stored securely, following local data privacy rules. 

\textbf{How to compensate?} One of the most important considerations for unbiased and ethical research is fair compensation of participants. Researchers should be aware of the minimum pay rates in the regions where the research is run, e.g., certain U.S. states have a higher rate than the national minimum. Even so, paying at the minimum might not encourage very active or eager participation. With the proliferation of remote work and schooling, we have seen increasing difficulties in motivating study participation. Under limited resources, % or limited participation enthusiasm, %researchers should be keenly aware that friend-sourcing and uncompensated volunteering can add bias to the results. Researchers should 
researchers can consider internally motivating participants (see also~\citet{cialdini2007influence}), for instance by gamifying the studies and providing participants with actionable feedback or personalized results~\citep{reinecke2015labinthewild}. 

\textbf{How to do this all safely and inclusively?} The COVID-19 pandemic has increased the amount of user research carried out virtually via online video conferencing software. Remote user research aids in disability and geographical inclusivity, since it is possible to include users from different regions and those who may not easily be able to travel to an in-person meeting. Virtual sessions allow participants to share their screen as they walk through previous projects and/or the proposed prototype, so the researcher can see exactly how they’re attempting to interact with the interface. Many videoconferencing options also enable efficient recording and transcription services. When conducting international research, one should be sensitive of timezones, compensation using local currency, and having translators on-hand if necessary. \citet{MediumInclusive} includes a full guide on running inclusive research with people with disabilities, who may require accommodations. 

% ---------------------------------------------------------

\section{Limitations and problems of user studies}\label{sec:limitations}

\zoya{In graphics and vision communities,} it is difficult to publish papers that do not include a table of scores that show better numbers for the proposed method compared to previous methods. But, for user evaluations, whether or not any of these numbers are meaningful is rarely scrutinized. \zoya{When the reviewing culture makes user studies a requirement for publication, without simultaneously providing standards for how to run them properly}\aaron{---or whether to run them at all---}it can harm research progress.
%In this section, we discuss ways in which user studies and user research---if performed poorly or used improperly---can harm research progress. %We first present reasons why current studies may not be yielding useful or accurate information. Second, we discuss how a culture of numerical evaluation–—in which it is difficult to publish a paper without a table of numerical scores—discourages innovative research. Third, we discuss the need for representative user samples, to avoid potentially biased or discriminatory findings. 
%For each topic, 
We make recommendations to help researchers, paper reviewers and conference chairs combat these effects.
% \zoya{There is a reviewing culture in graphics and vision encouraging user studies, or even making them a requirement for publication. However, these requests are not accompanied with standards about how to run these user studies. We conjecture that this may be contributing to perfunctory studies or studies that are not reproducible because not enough information or details are provided to reproduce them.}

\subsection{Replicability \& Reproducibility}

When we perform an experiment, such as testing the output of an algorithm, the results should be meaningful. Computer graphics and AI reviewing emphasize \textit{reproducibility} of the algorithm itself, e.g.,~\citet{bonneel2020code,pineau2021improving}. \textit{Reproducibility} means that the exact same experiment can be performed on the same data, e.g., when the authors release their data and code (\citep{NASEM}, p.~43). The results should also be generalizable, so the gold standard for experimental findings in mainstream science is \textit{replicability} (\citep{NASEM}, p.~43 based on~\citet{barba2018terminologies}). Findings are replicated when another researcher performs the same test on different data and/or different participants, and obtains the same result. For example, if another researcher were to compare your published algorithm to your baselines, on a new collection of users evaluating the results, would your algorithm perform just as well as when you tested it? If an experiment cannot be replicated, we generally treat the original findings as invalid. 
Note that definitions of reproducible/replicable are sometimes swapped~\citep{barba2018terminologies,plesser}.

%In our experience, many papers in graphics and vision do not provide enough information to rerun the same user study and reproduce the results. 
%More importantly, we conjecture that many user evaluations in graphics and vision papers might not replicate, driven by a research culture that incentivizes “user studies”, without standards for how to run them. 
%Reviewers often make including a “user study” a requirement for publication without attending to the study quality, thereby incentivizing perfunctory studies. 

%Should we be concerned? 
The Replication Crisis in mainstream science provides a cautionary tale \citep{open2015estimating}. 
Following these developments, many authors have raised replicability concerns for computer science~\citep{cockburn2020threats,peng2011reproducible}, specifically within machine learning~\citep{pineau2021improving,haibe2020transparency} and HCI~\citep{greenberg1991weak,greiffenhagen2014replication,hornbaek2014once,wilson2014replichi}. Many of the causes cited for these replication crises are relevant to computer graphics and vision, e.g., researchers can manipulate results–even unintentionally–such as by only reporting results from successful trials~\citep{simmons2011false}. 
%Of note is that output evaluations used in computer graphics and vision mirror similar studies in psychology for which replicability issues have been most prominent. 
%Given the casual nature of user evaluation in graphics and vision review processes, we expect similar replication problems. 

%On the flip side, user studies in computer graphics and vision may be replicable in cases when the outputs of proposed algorithms are visibly far better than the baselines, in which case the studies may be considered superfluous. 

\aaron{As a case study, \citet{WangIndeterminacy} performed several MTurk user studies testing versions of the same hypothesis, and got different results from different tests. While these studies were meant as exploratory, they illustrate how different studies may reach different conclusions on the same question.}

\subsection{Over-emphasis on user studies can hinder progress}\label{ssec:overemphasis}

Quantitative evaluations play an important role in computer science research to directly compare methods, and judge whether a new idea translates into improvements. 
However, numerical evaluations can be flawed for many reasons, and an overemphasis on imperfect metrics can distort research, as discussed in many prior contexts~\citep{biagioli2016watch,dehghani2021benchmark,stanley2015greatness,greenberg2008usability}. 

A good case study is an ICLR 2022 paper for which the reviews are available online \citep{einops}. Two reviewers gave very negative scores due, in part, to a lack of user studies. The paper was eventually accepted, accompanied by a summary chastizing the reviewers for using ``user studies'' as an excuse for poor reviewing, and accusing them of gatekeeping. 
%The full discussion is worth reading. 
The final decision noted that the submission described a software library that had been deployed for years, with thousands of users (information that was not revealed to the reviewers for anonymous review). Would the paper---which describes a highly impactful system---have been rejected if the committee had not had this information?  And, had the authors gone through the extra effort of contriving and performing a user study, would it have been meaningful, and would it have been enough to convince the reviewers? 

\textbf{Too-early quantification discourages innovation.} This emphasis on studies becomes harmful to the field when it discourages publication of innovative ideas that cannot easily be evaluated. In computer graphics research, we often seek to design new ways for users to create and design images, 3D, video, artwork, etc. For example, as pointed out by~\citet{greenberg2008usability}, in a searing critique of the over-emphasis on evaluation in HCI research, radical new interaction designs naturally will perform worse in many kinds of evaluations than polished mature systems. User research for early-stage designs (i.e., foundational research) must naturally be different from the numerical evaluation one might perform on a mature system.

\textbf{Reviewers and editors are responsible.} We have observed reviewers and editors impose onerous evaluation requirements on papers, without thought for whether such evaluations are meaningful or necessary. We have also observed authors and reviewers use \zoyagain{crowdsourced} evaluations as a crutch to avoid making hard decisions. Reviewer comments like ``I can’t tell if the images are better, maybe a user study would help'' are potentially harmful, encouraging authors to perform extra work that will not improve a lackluster paper.

\textbf{Recommendations.} We recommend that paper reviewers and area chairs take a critical eye toward evaluation: when requesting an additional user evaluation for a paper, to consider whether the evaluation would be meaningful or necessary for the paper.
Some of these problems can become systematic: the standards of what reviewers expect in a paper arise organically, as a result of authors’ own experience in previous papers, and what previous reviewers expected of them. Eventually these trends become common expectations, 
%If a culture of meaningless numerical evaluation comes to dominate, then this 
requiring a broader conversation in the field, and the efforts of program chairs to remind reviewers and area chairs to take the evidence and ideas in the paper as a whole, and not just look for the table of numbers.

\subsection{Bias and representation in study populations}\label{ssec:bias}

\citet{NobleAlgorithms} describes numerous examples where user research in product development led to biased systems because of failures to recruit diverse participants. Noble demonstrates that developers’ biases were not revealed in user studies because the studies performed included only similarly biased, white, cis-gender men as participants. Noble reveals deep-seated (and often unintentional) discrimination that has been baked into Google Image Search, Google PageRank, Airbnb rental offerings, ArtStor’s metadata, among others, resulting in biased outputs. In the context of graphics and vision work, it’s worth considering whether a system designed to reconstruct a user's hair would, for instance, work for different types of hair \citep{KimHair}. Further,~\citet{noble2016intersectional} discussed how algorithmic outputs are often evaluated according to pre-existing social norms and biases that include racism and sexism. 

\textbf{Recommendations.} Unfortunately, we’re not aware of any quick and easy solutions for approaching maximally inclusive research participant pools. There remain many open research questions, including which range of identities should be included in each study. A team might consider recruiting participants that exhibit a range of gender identities, ethnicities, sexual orientations, international locations and physical and mental abilities, depending on the technology being tested. 
For instance, when producing a tool that pertains to colors, it may be useful to include the perspective of someone who is colorblind. When constructing a novel interface, it will be helpful to include someone who uses a screenreader to determine its usability. For software that pertains to images of the human body (e.g., human hair or poses), it will be helpful to include a range of body types and ethnicities. For more tips on performing user research with participants with disabilities, see \citet{herman}.

\section{Conclusion}

User studies have occupied a contradictory place in computer graphics and vision research. On one hand, user studies are often considered a necessary step in the evaluation of research results. On the other hand, authors and reviewers often treat user studies as an afterthought, \zoya{not leaving sufficient time for the careful planning and reporting that replicable research requires.}
%sometimes thrown together in the final hours before a paper submission deadline, 
%and largely ignored by reviewers beyond checking that one set of numbers is less than another. This leads to a situation where considerable effort is wasted on performing user studies that are either trivial or not replicable.
%In this paper we have argued that, if the primary goal of running user studies is to appease reviewers rather than to generate new learnings, the utility and validity of such user studies should be put into question by authors and reviewers alike. 
Furthermore, penalizing work that does not contain user evaluation has the unintended consequence of incentivizing hastily done, poorly executed user research. A maxim to keep in mind is that \textbf{``bad user research leads to bad outcomes''}, and such research will continue if reviewers continue to ask for it. Our call to action for the computer graphics and vision communities is to put more thought into the requests for, design, and execution of user studies.

\zoya{On the other hand, when treated as a first-class citizen, commensurate with the algorithmic contributions of a paper,} user studies can provide valuable insights to help researchers iterate on and improve the design of their algorithms or interfaces. For instance, foundational research, including interviews and needfinding techniques, invoked at the start of the research and development process, offers the best ways to learn about users' wants and needs, even elucidating unexpected potential research directions. \zoya{For interfaces designed to support user workflows such as creative tasks, concept testing earlier prototypes or design concepts can inform future iterations, and usability studies can be invoked to evaluate later-stage prototypes. For evaluating algorithm outputs such as images or videos, qualitative feedback gathered alongside quantitative survey scores can help researchers understand how participants rate outputs, and which features are most important to them (which may be different from which features the researchers chose to optimize).  In all cases, we argue that user evaluations are more useful if the researchers can act on the insights gathered to improve the algorithm or tool, rather than viewing user studies as the step at the end of the research \& development road.}

\zoya{While crowdwork platforms have made certain types of user evaluation - particularly surveys - easier and faster to complete, they represent a narrow slither of the user research landscape. As an introduction to this broader landscape, and with an eye to what is most relevant to vision and graphics research, }
we covered a number of methodologies from user experience (UX), human-computer interaction (HCI), and related fields that can be invoked at various time points throughout the research \& development lifecycle. \zoya{We categorized user research into foundational research, interface evaluation, and output evaluation, and used the UX axes qualitative/quantitative, attitudinal/behavioral, and generative/evaluative when describing different methodologies. This additional precision in terminology helps elucidate the types of human data being captured in the evaluation.} 
For each of the categories of user research, we provided actionable guidelines and best practices, while referencing relevant resources for further reading. With this guidance in hand, we hope to contribute to improved quality standards for user studies in computer vision and graphics research.

% end of main matter

\section{Acknowledgements}
We thank Maneesh Agrawala, Wilson Chan, Valentin Deschaintre, Steven Diverdi, Jane E, Jose Echevarria, Matthew Fisher, Nick Kolkin, Sylvain Paris, Masha Shugrina, and Carol O'Sullivan for their valuable feedback on earlier drafts of this manuscript.
%\end{acknowledgements}

\bibliographystyle{ACM-Reference-Format}
\bibliography{references}

%\appendix

\end{document}